\documentclass[prd, twocolumn]{revtex4-1}
\usepackage{graphicx}
\usepackage[mathlines]{lineno}
\usepackage{amsmath}
\renewcommand{\vec}[1]{\mathbf{#1}}
\usepackage{slashed}
\usepackage{natbib}
\usepackage{bm}
\usepackage{newtxmath}
\usepackage{float}
\usepackage{amsfonts}
\usepackage{amssymb}
\usepackage{booktabs}
\usepackage{epsfig}
\usepackage{subfigure}
\usepackage{hyperref}
\hypersetup{
      colorlinks=true,
      citecolor=blue,
      linkcolor=blue,
      urlcolor=blue}
\everymath{\displaystyle}
\usepackage[usenames,dvipsnames]{color}
\usepackage{bm}
\newcommand{\bea}{\begin{eqnarray}}
\newcommand{\eea}{\end{eqnarray}}

\newcommand{\be}{\begin{equation}}
\newcommand{\ee}{\end{equation}}
\newcommand{\beq}{\begin{eqnarray}}
\newcommand{\eeq}{\end{eqnarray}}

\newcommand{\nn}{\nonumber}
\newcommand{\lt}{\left}     
\newcommand{\rt}{\right}
\newcommand{\grad}{\vec{\nabla}}
\newcommand{\vecB}{\vec{B}}
\def\bfig{\begin{figure}}
	\def\efig{\end{figure}}
\newcommand{\bsym}{\boldsymbol}
\newcommand{\bk}{{\bsym k}}
\newcommand{\calA}{{\cal A}}
\newcommand{\calH}{{\cal H}}

\newcommand{\Bt}{\tilde{B}}
\def\eqn#1{eq.\ (\ref{#1})}

\begin{document}
\title{Chiral plasma instability and primordial Gravitational wave}
\author{Sampurn Anand$^{2}$}
  \email{sampurn@prl.res.in}
\author{Jitesh R. Bhatt$^{1}$}
  \email[e-mail:]{jeet@prl.res.in}
\author{Arun Kumar Pandey$^{1}$}
  \email[e-mail:]{arunp@prl.res.in}
\affiliation{%
		\centerline{Physical Research Laboratory, Theoretical Physics Division, Ahmedabad 380 009, India$^{1}$}
		\centerline{St. Stephen's College, University Enclave, New Delhi- 110007, India$^2$}
	        }
%
\begin{abstract}
	It is known that cosmic magnetic field, if present, can generate anisotropic stress in the plasma and hence, can act as a source of gravitational waves. These cosmic magnetic fields can be generated at very high temperature, much above electroweak scale, due to the gravitational anomaly in presence of the chiral asymmetry. The chiral asymmetry leads to instability in the plasma which ultimately leads to the generation of magnetic fields. In this article, we discuss the generation of gravitational waves, during the period of instability, in the chiral plasma sourced by the magnetic field created due to the gravitational anomaly. We have shown that such gravitational wave will have a unique spectrum. Moreover, depending on the temperature of the universe at the time of its generation, such gravitational waves can have wide range of frequencies. We also estimate the amplitude and frequency of the gravitational waves and delineate the possibility of its detection by the future experiments like eLISA. 
\end{abstract}
\maketitle
\section{Introduction}
\label{sec:intro}
Gravitational wave (GW) once generated, propagates almost unhindered through the space-time. This property makes GW a very powerful probe of the source which produces it as well as the medium through which it propagates (see \cite{Maggiore:2006uy, Buonanno:2007yg, Hogan:2006va} and references therein). From the cosmological point of view, the most interesting gravitational radiation is that of the stochastic gravitational wave (SGW) background. Such gravitational radiations are produced by events in the early stages of the Universe and hence, may decipher the physics of those epochs. Several attempts have already been made in this regard and various sources of SGW have been considered. List of SGW source includes quantum fluctuations during inflation
~\cite{Rubakov:1982df, BRatra:1992br, Giovannini:1999m, Sahni:1990tx}, bubble wall collision during phase transition ~\cite{Kamionkowski:1994km, Kosowsky:1992tw, 
	Kosowsky:1994tm, Riccardo:2001an, Witten:1984we, hogan:1986cj}, cosmological magnetic fields ~\cite{Deriagin:1987gr, Giovannini2000ms, Kosowsky:2002tm, Sidorenko:2016vom} and turbulence in the plasma~\cite{Kosowsky:2002tm, Dolgov:2002gn, Nicolas:2004an}.

In the early universe, before the electroweak phase transition, many interesting phenomena have taken place. For instance, it has been shown by several authors~\cite{joyce:1997sh, Giovannini1998sha, Boyarsky:2012a, Tashiro:2012mf, Pandey:2015kaa, Bhatt:2015ewa} that in presence of asymmetry in the left-handed and the right-handed particles in the early Universe, there will be an instability which leads to the generation of turbulence in the plasma as well as (hyper-charge) magnetic fields. In reference \cite{Anand:2017zpg}, it has been shown that these magnetic fields can be generated even in absence of net chiral charge but due to the gravitational anomaly. The magnetic field generated via this mechanism are helical in nature. However, helicity ($\mathcal{H}_B =\frac{1}{V} \int d^3 x\, \, {\bf A} \cdot {\bf B}$) of these magnetic fields are not completely conserved due to the fact that large but finite conductivity gives a slight time variation of the helicity density. In presence of chiral imbalance, the chiral charge conservation equation, which is valid at temperature $T>T_R \approx 80$ TeV~\cite{joyce:1997sh}, is given as:
$\partial_\eta (\Delta \mu + \frac{\alpha'}{\pi} \mathcal{H}_B) = 0$
where $\Delta \mu=\mu_R - \mu_L$, $\alpha'$ and $\eta$ are the asymmetry in the chiral chemical potentials and $U(1)_Y$ fine structure constant and conformal time respectively.
At the onset of the instability, $\mathcal{H}_B \approx 0$ and at subsequent time, helicity will grow at the expense of chiral chemical potential. In this regime, $\Delta \mu$ can be regarded as constant. On the other hand, 
at temperature $T< 80$ TeV the above conservation equation is not valid as chiral flipping $\Gamma_{f}$ rate is non-vanishing and the hence, the conservation equation is given by~\cite{Boyarsky:2012a}
\begin{equation}
\frac{d}{d \eta}(\Delta \mu + \frac{\alpha'}{\pi} \mathcal{H}_B)=-\Gamma_f \Delta \mu
\label{eq:delmucons}.
\end{equation}
Here $ \Gamma_f=\frac{T_R}{M_*}T$ and $M_* = \lt({90\over 8\pi^3g_{{\rm eff}}}\rt)^{1/2}M_{pl}\,$, where $g_{{\rm eff}}$ and $M_{pl}=1/\sqrt{G}$ are relativistic degree of freedom and the Planck mass respectively \cite{joyce:1997sh}. In this regime, non-linearity sets in and the magnetic fields are generated. The generated magnetic fields show inverse cascade behaviour, where magnetic energy is transfered from small scale to large-scale. In reference \cite{Fujita:2016kk}, it is shown that the currently observed baryon asymmetry ($\eta_B \sim 10^{-10}$) can be generated if the magnetic fields produced above electroweak scale undergoes the inverse cascade and the strength of the magnetic field is of the order $(10^{-14}-10^{-12})$ G at the galactic scale. The generated magnetic fields induce a anisotropic stress so that their energy density $B^2/8\pi$ must be a small perturbation, in order to preserve the isotropy of the Friedmann-Robertson-Walker background. This condition allows us to use the linear perturbation theory. In this perturbation scheme, peculiar velocity and magnetic fields are considered to be first order in perturbations. At sufficiently large length scale, the effect of the fluid on the evolution of magnetic fields can be neglected. However, at small length scales the interaction between the fluid and the magnetic field become very crucial. At the intermediate length scale, plasma undergoes Alfven oscillations and on a very small scale (viscous scale), these fields undergo exponential damping due to the shear viscosity \cite{Caprini:2004cdr}. Thus, the large-scale magnetic fields are important for the physics at the cosmic scale.
We have already mentioned that the seed magnetic field can be generated even in absence of net chiral charge but due to gravitational anomaly \cite{Anand:2017zpg} and the magnetic field thus generated can produce instability in the plasma. These magnetic fields contribute a anisotropic stress to the energy-momentum tensor and hence can act as a source for the generation of the GWs. The underlying physics of GW generation is completely different from previously considered scenarios. Therefore, it is important to investigate the generation and evolution of GW in this context. In this article, we compute the metric tensor perturbation due to the chiral magnetic field. Since chiral magnetic field, which sources the tensor perturbations, has a unique spectrum, the GWs generated is expected to have a unique signature in its spectrum as well. Moreover, we compute the amplitude and frequency of the GW and show its dependences on the model parameters. Consequently, any detection of SGW in future measurements like eLISA 
will constrain or rule out such theoretical constructs.

This paper is organized as follows: in section (\ref{sec:ch_mag_field}) we outline the generation and
evolution of magnetic field due to gravitational anomaly and chiral imbalance. We discuss the 
generation of SGW in section (\ref{sec:GW_SMF}). We present our results in section (\ref{sec:results}) and finally conclude in ~(\ref{sec:conc}).
Throughout this work, we have used $\hslash=c=k_B=1$ unit. We have also considered Friedman-Robertson-Walker metric for expanding background space-time
\begin{equation}
ds^2=a^2(\eta)\lt(-d\eta^2+\delta_{ij}\,dx^i\,dx^j\rt)\, ,
\label{eq:FLRW-backGround}
\end{equation}
where scale factor $a(\eta)$ have dimension of length, whereas conformal time $\eta$ and conformal coordinate $x^i$ are dimensionless quantities. In the radiation dominated epoch $a=1/T$, we can define conformal time $\eta = M_*/T$. Unless stated otherwise, we will work in terms of comoving variables defined as,
\begin{eqnarray}
B_c = a^2(\eta)B(\eta),~~
\mu_c = a(\eta)\mu,~~ k_c =a\, k, 
\label{eq:com-varia}
\end{eqnarray}
where, $B$, $\mu$ and $k$ represents the physical magnetic fields, chemical potential, and the wave number respectively. It is clear from the convention used here that all the comoving quantities are dimensionless. In terms of these comoving variables, the evolution equations of fluid and electromagnetic fields are form invariant  \cite{Holcomb:1989tf, Gailis:1995ohk, Dettmann:1993zz}. Therefore, we will work with the above defined comoving quantities and omit the subscript ``c'' in our further discussion.
\section{Gravitational anomaly and magnetic fields in the early universe}
\label{sec:ch_mag_field}
Although the origin of large-scale magnetic field is still an unsolved issue in cosmology, several attempts have been made to address the issue. It has been discussed in the literature that there are processes in the early universe, much above electroweak scale, which can lead to more number of right-handed particle than the left-handed ones and remains in thermal equilibrium via its coupling with the hypercharge gauge bosons ~\cite{Kuzmin:1985mm,tHooft1976az}. Furthermore, 
if the plasma has rotational flow or external gauge field present, there could be a current in the direction parallel to the vorticity due to rotational flow or parallel to the external field. The current 
parallel to the vorticity is known as chiral vortical current and the phenomenon is called as chiral vortical effects (CVE) ~\cite{PhysRevD.20.1807,Johanna:2009,Banerjee2011,son:2009,Landsteiner:2011}. Similarly, the current parallel to the external magnetic field is 
known as chiral magnetic current and the phenomenon is called as chiral magnetic effect (CME) \cite{Fukushima:2008xe, PhysRevD.22.3080, Neiman2011, NIELSEN1983389, Alekseev:1998ds}. CVE and CME are 
characterized by the transport coefficients $\xi$ and $\xi^{(B)}$ respectively. The form of these coefficients can be obtained by demanding the consistency with the second law of thermodynamics ($\partial_\mu\,s^\mu \geq 0$, with $s^\mu$ being the entropy density). Thus, in presence of chiral imbalance and gravitational anomaly, which arises due to the coupling of spin with gravity~\cite{AlvarezGaume:1983ig}, the coefficients for each right and left particle have the following form \cite{son:2009,Landsteiner:2011,Neiman2011}
\begin{equation}
\xi_i = C\,\mu_i^2\,\lt[1-\frac{2\,n_i\,\mu_i}{3\,(\rho + p)}\rt]\, +
\frac{D\,T^2}{2}\lt[1-\frac{2\,n_i\,\mu_i}{(\rho + p)}\rt] \, ,
\label{eq:xi} 
\end{equation}
\begin{equation}
\xi_i^{(B)} = C\,\mu_i\,\lt[1-\frac{n_i\,\mu_i}{2\,(\rho + p)}\rt]\, -
\frac{D}{2}\lt[\frac{n_i\,T^2}{(\rho + p)}\rt]\, .
\label{eq:xiB}
\end{equation}
In above equations `$i$' stands for each species of the chiral plasma. The constants $C$ and $D$ are related to those of the chiral anomaly and mixed gauge-gravitational anomaly and are given as $C=\pm 1/4\pi^2$ and $D=\pm1/12$ for right and left-handed chiral particles respectively. The variables $n$, $\rho$ and $p$ are respectively the number density, energy density and pressure density.

Using the effective Lagrangian for the standard model, one can derive the generalized Maxwell's eq. \cite{Semikoz:2011tm},  
\begin{equation}
\grad \times \vecB = \vec j\, ,
\label{eq:GenMax}
\end{equation}
where ${\vec j}$ is defined as ${\vec j} = {\vec j}_{\rm v} + {\vec j}_5$ with ${\vec j}_{\rm v}$ being the vector current and ${\vec j}_5 $ is the axial current. Vector and axial currents respectively takes the following form:
\begin{eqnarray}
j^\mu_{\rm v} & = & n_{\rm v}\,u^\mu  + \sigma E^\mu + \xi_{\rm v}\,\omega^\mu +
\xi_{\rm v}^{(B)}\,B^\mu\, ,
\label{eq:VectorCurrent} \\
j_5^\mu & = & n_5\,u^\mu 
+ \xi_{\rm 5}\,\omega^\mu + \xi_{5}^{(B)}B^\mu \, .
\label{eq:AxialCurrent}
\end{eqnarray}
In the above equations, any quantity $x_{\rm v, (5)}$ denotes the sum (difference) of the quantities pertaining to right and left handed particles. Also $E^\mu=u_{\nu} F^{\mu\nu}$, $B^\mu=1/2 \varepsilon^{\mu\nu\sigma\delta} u_\nu F_{\sigma\delta}$, and $\omega^\mu= 1/2 \varepsilon^{\mu\nu\sigma\delta} u_\nu\partial_\sigma u_\delta$ are the electric, magnetic and the vorticity four vectors respectively. We have ignored the displacement current in eq.(\ref{eq:GenMax}). Taking $u^\mu = (1, {\bf v})$ 
and using eq.(\ref{eq:VectorCurrent})- eq.(\ref{eq:AxialCurrent}), one can show
that
\begin{eqnarray}
j^0   & = & n = n_{\rm v} + n_5 
\label{eq:totCurrent1} \\
\vec{j}  & = & n\vec{\rm v} + \sigma(\vec{E} +\vec{\rm v}\times \vec{B})  +
\xi \, \vec{\omega} + \xi^{(B)}\,\vec{B}  \label{eq:totCurrent2}\, ,
\end{eqnarray}
with $\xi = \xi_{\rm v} + \xi_5$ and $\xi^{(B)} = \xi^{(B)}_{\rm v} + \xi^{(B)}_5$.  Assuming the velocity field to be divergence free field, {\it i.e.} $\grad\cdot\vec{\rm v} = 0$ and taking curl of eq.(\ref{eq:GenMax}) along with the expression for current from eq.(\ref{eq:totCurrent2}) we obtain,
\begin{eqnarray}
{\small \frac{\partial \vecB}{\partial \eta} =  \frac{n }{\sigma}\boldsymbol{\omega} +\frac{1}{\sigma}\nabla^2\vecB + \grad\times(\vec{\rm v}\times \vecB) + \frac{\xi }{\sigma}\grad\times \boldsymbol{\omega}
	+  \frac{\xi^{(B)}}{\sigma}\grad\times\vecB.}
\label{eq:magevol}
\end{eqnarray}
In our previous work \cite{Anand:2017zpg}, we discussed that the seed magnetic field (for which $\vecB$ in the right hand side of \eqn{eq:magevol} is zero) can be generated even if $n = 0$. The $T^2$ term in $\xi_i$ (see \eqn{eq:xi}), which arises due to the gravitational anomaly, acts as a source for the generation of seed field. On the other hand, presence of finite chiral imbalance such that $\mu/T \ll 1$, $T^2$ term in $\xi$ still acts as source of seed magnetic field but non-zero $\xi^{(B)}$ triggers instability in the system. This result is in agreement with the previous studies where it was shown that in the presence of a chiral imbalance in the plasma, much above the Electroweak scale ($T>100~{\rm GeV}$), there can be instability known as chiral plasma instability \cite{Akamatsu:2013pjd}.

The production and evolution of the magnetic field can be seen through the evolution equation given in \eqn{eq:magevol}. In order to do so, we decompose the divergence-free vector fields, {\it e.g.} magnetic 
field, in the orthonormal helicity basis, $\varepsilon^{\pm}_i$, defined as
\be
\bsym{ \varepsilon}^\pm({\bf k}) = \frac{-i}{\sqrt{2}}\lt[{\bf e}_1({\bf k})\,\pm \,i\, {\bf e}_2({\bf k})\rt]\exp(i \, {\bf k} \cdot{\bf x}),
\ee
where $({\bf e}_1, {\bf e}_2, {\bf e}_3 = \hat {\bf k})$ form a right-handed orthonormal basis with ${\bf e}_2 =  {\bf \hat k} \times {\bf e}_1$. We choose ${\bf e_1} $ to remain invariant under the transformation ${\bf k}\rightarrow -{\bf k}$ while ${\bf e_2} $ flip its sign. In this basis, the magnetic field can be decomposed as
\begin{equation}
B_i(\eta, {\bf k})= B^+(\eta, {\bf k})\,\varepsilon_i^+({\bf k})\, + \, 
B^-(\eta, {\bf k})\, \varepsilon_i^-({\bf k})\, .
\end{equation}
In this basis, the evolution equation for the $|B^{\pm}|^2$ can  be obtained from \eqn{eq:magevol}
\begin{eqnarray}
\frac{\partial |\Bt^\pm|^2}{\partial \eta}  =  \frac{2}{\sigma}\lt(-k^2\pm \xi^{(B)} k\rt)&  |\Bt^\pm|^2
+ \frac{2}{\sigma^2}\lt(\pm nk + \xi k^2\rt)^2 \nonumber \\
& \times |{\rm v}^\pm|^2{\mathcal F}_s
\label{eq:b-modes}
\end{eqnarray}
where ${\mathcal F}_s=\eta-\eta_0$ for $\eta-\eta_0\leq 2\pi/(k{\rm v})$ and zero for $\eta-\eta_0\geq 2\pi/(k{\rm v})$. Also magnitude of the wave vector ${\bf k}$ is represented as $k$, i.e. $|{\bf k}|=k$.~
From eq. (\ref{eq:b-modes}), it is clear that, when first term is dominant over the second term, magnetic modes will grow exponentially with time as\\
\begin{equation}
|B^\pm|^2 = |B_0|^2 \exp(\frac{2\eta}{\sigma}k_{\rm ins}^2)\exp(-\frac{2\eta}{\sigma}(k\mp k_{\rm ins})^2) \nonumber\, ,
\end{equation}
where $k_{\rm ins}= \alpha' \Delta\mu/\pi$ is the value of the $k$ at which magnetic modes have maximum growth rate. The exponential growth of the magnetic modes is true only in the linear regime. In this regime, chiral chemical potential remain constant and magnetic fields are generated at the cost of chiral imbalance. However, when magnetic field generated is sufficiently large, the non-linear effects become prominent. In this case, we need to consider the evolution of the chemical potential~\cite{Tashiro:2012mf, joyce:1997sh}, which is given by eq. (\ref{eq:delmucons}). At temperature $T> 80$ TeV, perturbative processes that lead to the flipping of the chirality are small compared to the expansion rate of the Universe and hence we can consider the chiral symmetry to be an exact symmetry of the theory. Therefore, a chemical potential $\mu_{R, L}$ for each species can be defined. However, at $T< 80$ TeV, the chiral symmetry is not exact due to the chiral flipping of the right-handed particles to the left-handed particles and vice versa. As a consequence, the number densities of these particle are not  conserved and therefore, we can not define the chemical potentials for right and left handed particles \cite{Boyarsky:2012a, Boyarsky:2012as}. In order to obtain the velocity profile, we used the scaling symmetry ~\cite{Olesen:1996ts,Yamamoto:2016xtu} rather than solving the Navier- Stokes equation. In our earlier work \cite{Anand:2017zpg}, we have obtained $|{\rm v}^+|=|{\rm v}^-| = \pi k^{-3/2}{\rm v}$. Here ${\rm v}$ is given by
\begin{eqnarray}
{\rm v}(k, \eta)= {\rm v}_i \left(\frac{k}{k_i(\eta_0)}\right)^{(1+m)/2}\left(\frac{\eta}{\eta_0}\right)^{(m-2)/6}.
\end{eqnarray}
In above equation, $m$ is a positive integer and ${\rm v}_i$, $k_i$ are arbitrary function encoding the boundary conditions. We also showed that the scaling law allows more power in the magnetic field. For $n = 0$, $E_B\propto k^7$ at larger length scale  whereas for $n\neq0$, $E_B\propto k^5$ instead of $k^7$~\cite{Tashiro:2012mf}. However, in both the scenarios $E_B$ is more than that of the case without considering the scaling symmetry ~\cite{Tashiro:2012mf}. This aspect is important for this analysis as more power in the helical magnetic field can generate larger anisotropic stress. 

The two point correlation function for the helical magnetic fields is given as:
\begin{eqnarray}
\langle B_i({\bk)}B^*_j({\bk'})\rangle ={(2\pi)^3 \over 2} 
\delta({\bk-\bk'}) \lt[ P_{ij}S(k)+i\,\epsilon_{ijl}\hat k_l\calA (k)\rt]~~
\label{eq:B-B-corel}
\end{eqnarray}
where $S(k)$ is the symmetric and $\calA(k)$ is the helical part of the magnetic field power spectrum. $P_{ij} = \delta_{ij} - \hat k_i\,\hat k_j$ is the transverse plane projector which satisfies: $P_{ij}\hat k_j = 0 $, $P_{ij}P_{jk} = P_{ik}$ with $\hat k_i = k_i/k$ and $\epsilon_{ijk}$  is the totally antisymmetric tensor. Using  \eqn{eq:B-B-corel} and the reality condition $B^{\pm*}({\bsym k})=-B({\bsym k})^{\pm}$, one can show that
\begin{eqnarray}
\langle B^+({\bsym k})B^{+*}({\bsym k'}) + B^-({\bsym k})B^{-*}({\bsym k'})\rangle 
& = & -(2\,\pi)^3 S(k)\delta({\bsym k -\bsym k'})\\
\langle B^+({\bsym k})B^{+*}({\bsym k'}) - B^-({\bsym k})B^{-*}({\bsym k'})\rangle 
& = &  (2\,\pi)^3\,\calA(k)\delta({\bsym k - \bsym k'}).
\end{eqnarray}
Note that, $\calA({\bk})$ represents the difference in the power of left-handed and right-handed magnetic fields, however a maximally helical magnetic fields  
configuration can be achieved when $\calA({\bk})=S({\bk})$. 
\subsection{Anisotropic stress}
Tensor component of metric perturbation, which results in the gravitational waves, are sourced by the transverse-traceless part of the stress-energy tensor. In this work, we assume that the anisotropic stress is generated by the magnetic stress-energy tensor which is given by
\begin{equation}
T_{ij}= \frac{1}{a^2}\lt( B_i\,B_j-\frac{1}{2}\delta_{ij} \, B^2\rt)\, ,
\end{equation}
Note that the spatial indices are raised, lowered and contracted by 
the Kronecker delta such that $B^2 = \delta^{ij}B_iB_j$. The magnetic field component $B_i$ then coincide with the 
comoving magnetic field which in our notation is $B_c = a^2\, B$. In Fourier space, the stress-energy tensor for the magnetic field take the following form 
\begin{equation}
\small{T_{ij}({\bsym k})=\frac{a^{-2}}{2(2\pi)^4}\int d^3q 
	\lt[
	B_i(\bsym q)B^*_j({\bsym q}-{\bsym k}) - 
	\frac{1}{2}B_l({\bsym q})B_l^*({\bsym q}-{\bsym k})\delta_{ij}
	\rt]}.
\label{eq:Fou-mag-stress}
\end{equation}
We are interested in the generation of GW and hence, we need to extract the transverse traceless component of the stress energy tensor given in~\eqn{eq:Fou-mag-stress}. This can be done by using the projection operator:
$\mathcal{P}_{iljm}({\bk})=[P_{il}({\bk})\, {P}_{ jm}({\bk})-\frac{1}{2} \,P_{ij}({\bk})\,P_{lm}({\bk})]$
which leads to~\cite{Durrer:1998rm}
\begin{eqnarray}
\Pi_{ij}(\bk) = \mathcal{P}_{iljm}({\bk})\, T_{lm}(k) .
\end{eqnarray}
At this stage, we will evaluate the two-point correlation function of the energy-momentum tensor which will be used in the later part of the calculation. The two point correlation of the stress-energy tensor takes the form
\begin{equation}
\begin{split}
\langle T_{ij}(\bk)T_{lm}^*(\bk') \rangle = &{1 \over (2\pi)^6}{1\over (4\pi)^2 a^4}\int d^3 p\int d^3 q  \\ 
&\bigg[\langle 
B_i(\bsym p)\,B_j^*(\bsym p-\bk)\,B_l^*(\bsym q) B_m(\bsym q -\bk')
\rangle  \\ &+ (..)\delta_{ij} + (..)\, \delta_{lm} + (..)\delta_{ij} \delta_{lm}\bigg]
\end{split}
\end{equation}
It was shown in ref.~\cite{Mack:2001gc} that only first term in the angular bracket will have a non-vanishing contribution in the two-point correlation function of the anisotropic stress $\langle\Pi_{ij}\Pi^*_{lm}\rangle$. Therefore, we will not evaluate other terms. Moreover, around the chiral instability, the magnetic field profile is 
Gaussian and the major contribution to the anisotropic stress come from this regime only. Therefore, we can safely
assume that the magnetic fields are Gaussian and hence four-point correlator in the integrand can be expressed, using Wick's theorem, in terms of two-point correlators as
\be
\begin{split}
	\langle B_i(\bk_i)B_j(\bk_j)B_l(\bk_l)&B_m(\bk_m)\rangle = \\
	&\langle B_i(\bk_i) B_j(\bk_j)\rangle\,\langle B_l(\bk_l) B_m(\bk_m)\rangle \\
	&+  \langle B_i(\bk_i) B_l(\bk_l)\rangle\,\langle B_j(\bk_j) B_m(\bk_m)\rangle \\
	&+ \langle B_i(\bk_i) B_m(\bk_m)\rangle\,\langle B_j(\bk_j) B_l(\bk_l)\rangle\,.
\end{split}
\ee
After a bit of lengthy but straightforward calculation, we obtain the two point correlations of the energy momentum tensor and the transverse-traceless part of the energy momentum tensor is
\be
\begin{split}
	\langle {\rm T}_{{\rm ij}}(\bsym k){\rm T}_{{\rm lm}}^*(\bsym k') \rangle &={1\over 4 (4\pi)^2 a^4}\delta(\bk -\bk')\int d^3 p \bigg[ S(p)S(k-p) \\ 
	&\lt\{ P_{il}(\hat{\bsym p})\,  P_{jm}\widehat{(\bsym k - \bsym p)}
	+ P_{im}(\hat{\bsym p})\,  P_{jl}\widehat{(\bsym k - \bsym p)}\rt\} \\
	&  - {\cal H}(p)\,{\cal H}(k-p)\big\{\epsilon_{ila}\,\epsilon_{jmb }\,\hat{\bsym p}_a\,\widehat{({\bsym k} - \bsym p)}_b \\
	& + \epsilon_{imc}\,\epsilon_{jld }\,\hat{\bsym p}_c\,\widehat{(\bk - \bsym p)}_d\big\} 
	+ i\, {\cal H}(p)S(k-p) \\
	&\big\{\epsilon_{ila}\,P_{jm}(\widehat{\bsym k - \bsym p})\,\hat{\bsym p}_a\, 
	+ \epsilon_{imc}\,P_{jl}(\widehat{\bsym k - \bsym p})\,\hat{\bsym p}_c\,\big \} \\
	& + i\, {\cal H}(k-p)S(p)\big\{\epsilon_{jmb}\,P_{il}(\hat{\bsym p})\, (\widehat{\bsym k - \bsym p})_b\\
	&\, + \epsilon_{jld}P_{im}(\hat{\bsym p})\,(\widehat{\bsym k - \bsym p})_d\,\big\} 
	\bigg]
	\label{eq:stress-corel}
\end{split}
\ee
and
\be
\langle \Pi_{{\rm ab}}({\bk})\,\Pi_{{\rm cd}}^*({\bk}') \rangle = \mathcal{P}_{{\rm aibj}}({\bk})\,\mathcal{P}_{{\rm cldm}}({\bk}')\langle {\rm T}_{{\rm ij}}({\bk})\,{\rm T}_{{\rm lm}}^*({\bk}') \rangle .
\label{pi-correlation1}
\ee
Above equation can also be written in terms of a most general isotropic transverse traceless fourth rank tensor which obeys ${\cal P}_{abcd}={\cal P}_{bacd}={\cal P}_{abdc}={\cal P}_{cdab}$ as~\cite{Caprini:2001cdr, Caprini:2004cdr}
\be
\langle \Pi_{{\rm ab}}({\bsym k})\Pi_{{\rm cd}}^*({\bsym k}') \rangle = \frac{1}{4a^4}[\mathcal{M}_{abcd}f(k)+i\mathcal{A}_{abcd} g(k)]\delta({\bsym k}-{\bsym k}'),
\ee
with a definition of $\mathcal{M}_{abcd}$ and $\mathcal{A}_{abcd}$ as
\begin{eqnarray}
\mathcal{M}_{abcd} &=& P_{ac}P_{bd}+P_{ad}P_{bc}-P_{ab}P_{cd}  \,\\
\mathcal{A}_{abcd} &=& \frac{1}{2}\hat{{\bsym k}}_e (P_{bd}\epsilon_{ace}+P_{ac}\epsilon_{bde}+P_{ad}\epsilon_{bce}+P_{bc}\epsilon_{ade}) \nonumber
\end{eqnarray}
which follows following properties:  
\beq
\mathcal{M}_{abcd} & = & \mathcal{M}_{cdab}= \mathcal{M}_{abdc}=\mathcal{M}_{bacd}\nn\,\\
\mathcal{A}_{abcd} & = & \mathcal{A}_{cdab}= -\mathcal{A}_{abdc}=-\mathcal{A}_{bacd}\nn\,\\
\mathcal{M}_{abab} & = & 4\nn\,\\
\mathcal{M}_{aacd} & = & \mathcal{M}_{abcc} = 0\nn\,\\
P_{ea}\mathcal{M}_{abcd} & = & \mathcal{M}_{ebcd} \nn\, \\
P_{ea}\mathcal{A}_{abcd} & = & \mathcal{A}_{ebcd} \nn\,\\
\mathcal{M}_{abcd}\,\mathcal{M}_{abcd} & = & \mathcal{M}_{abcd}\,\mathcal{M}_{abcd} =8\nn\, \\
\mathcal{A}_{abcd}\,\mathcal{M}_{abcd} & = & 0 \nn\,\\
\mathcal{A}_{abab} = \mathcal{A}_{aacd} & = & \mathcal{A}_{abcc}=0.
\eeq
The functions $f(k)$ and $g(k)$ are defined as follows:
\beq
\delta({\bk}-{\bk}')\, f(k) & =& \frac{1}{2}\mathcal{M}_{abcd}\langle T_{ab}({\bk})\,T_{cd}^*({\bk}')\rangle \nn\, \\
\delta({\bk}-{\bk}')\, g(k) & =& -\frac{i}{2}\mathcal{A}_{abcd}\langle T_{ab}({\bk})\,T_{cd}^*({\bk}')\rangle .
\label{f-g-function}
\eeq
We point out that the functions $f(k)$ and $g(k)$ also depends on time because 
of the time dependence of the magnetic fields. The integral form of these two functions are
\be
\begin{split}
	f(k) = \frac{1}{4}\frac{1}{(4\pi)^2} \int d^3p\,\big[&(1+\gamma^2)(1+\beta^2)\,S(p)S(k-p) \\
	&+ 4\,\gamma\,\beta\,\calH(p)\calH(k-p) \, \big]
\end{split}
\ee
\be
g(k) = \frac{1}{2}\frac{1}{(4\pi)^2}\int d^3p\, \lt[(1+\gamma^2)\,\beta\,S(p)\, \calH(k-p)\rt].
\ee
where $p=|{\bf p}|$, $(k-p)=|{\bk}-{\bf p}|$, $\gamma= \hat{{\bk}}\cdot \hat{{\bf p}}$ and $\beta= \hat{{\bk}}\cdot(\widehat{{\bk}-{\bf p}})= (k-p\gamma)/\sqrt{k^2+p^2-2\gamma p k}$. 
\section{Gravitational waves from chiral magnetic fields}
\label{sec:GW_SMF}
\begin{figure}[h!]
	\includegraphics[width=\columnwidth]{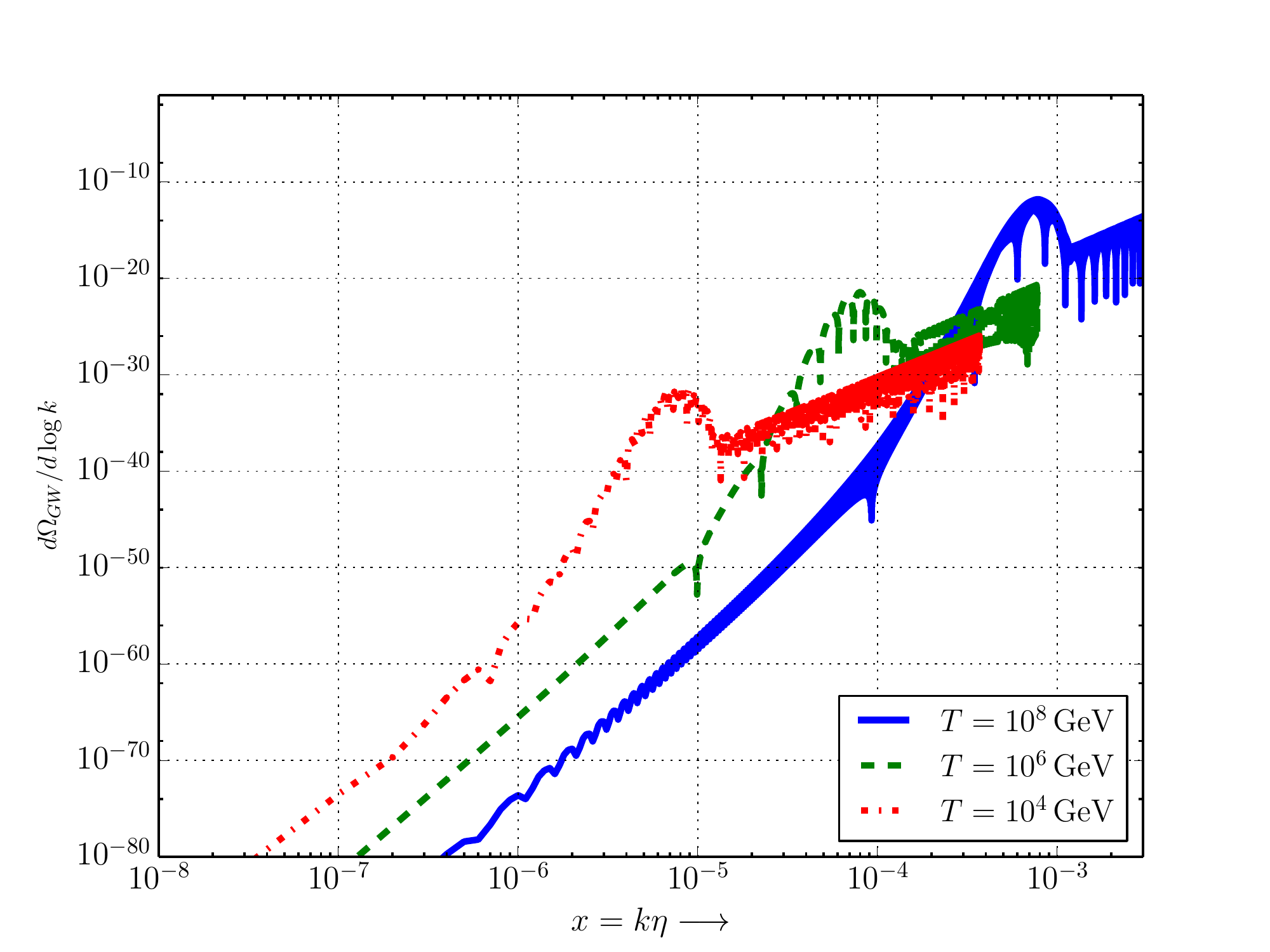}
	\caption{Gravitational wave spectrum as a function of $x=k\eta$. We have fixed $n = 10^{-6}$ and varied temperature.}
	\label{fig:chgw_k_diff_T}
\end{figure}

\begin{figure}[h!]
	\includegraphics[width=\columnwidth]{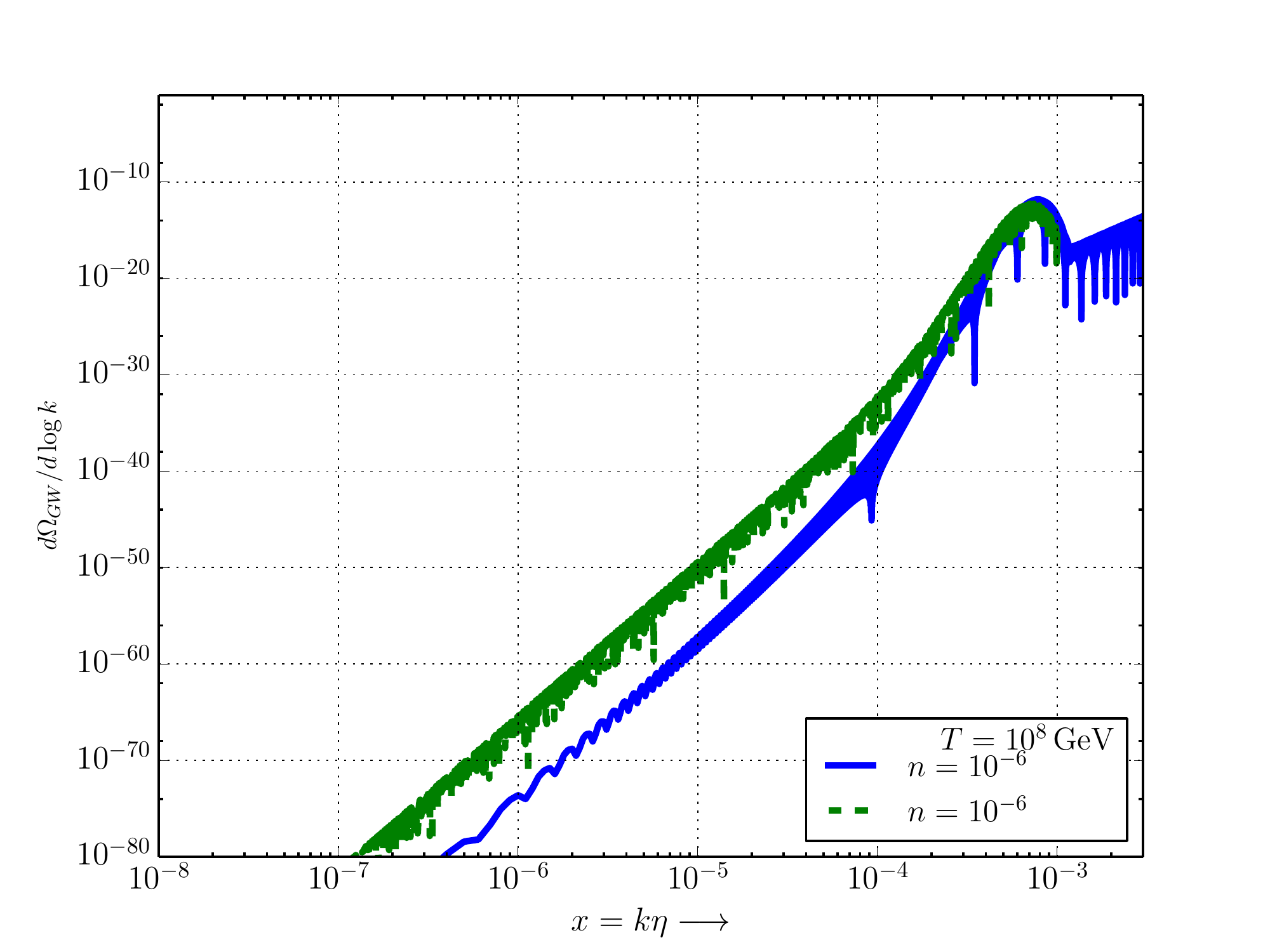}
	\caption{Gravitational wave spectrum as a function of $x=k\eta$ at different number densities at temperature T$\sim 10^8$~GeV. For a large number density, the effects at large length scale is more than at the small length scale. }
	\label{fig:chgw_k_diff_N}
\end{figure}

We have seen that the chiral magnetic field generated at very high temperature can produce anisotropic stress which leads to tensor perturbation in the metric. To linear order, the small tensor perturbation in the FLRW background can be written as:
\be
ds^2= a^2(\eta)[-d\eta^2+(\delta_{ij}+2 h_{ij})dx^i\, dx^j],
\ee
where the tensor perturbation satisfies the following conditions $h_i^i=0$ and $\partial^i \, h^j_i=0$. In this gauge, these tensor perturbations describe the GW whose evolution equation can be obtained by solving Einstein's equation which, to the linear order in $h_{ij}$, is given as:
\begin{eqnarray}
h_{ij}'' + 2\,H\,  h_{ij}' \, +k^2\, h_{ij}= 8\,\pi\, G \, \Pi_{ij},
\label{eq:tensor-ptbn}
\end{eqnarray}
where prime denotes the derivative with respect to the conformal time $\eta$ and $H =\frac{1}{a(\eta)} \frac{\partial a(\eta)}{\partial \eta}$. The time dependence in the right hand side of the \eqn{eq:tensor-ptbn} comes from the fact that the magnetic field is frozen in the
plasma. Therefore, $\Pi_{ij}(k,\eta)$ is a coherent source, in the sense that each mode 
undergoes the same time evolution. Assuming that the tensor perturbations has a Gaussian 
distribution function, the statistical properties can be described by the two-point 
correlation function given as,
\be
\begin{split}
	\langle h_{ij}^{*'}({\bsym k}, \eta)\, h_{lm}'({\bsym k}', \eta)\rangle = \frac{1}{4}
	\delta^3({\bsym k}-{\bsym k}')&\big[\mathcal{M}_{ijlm} S_{GW}(k,\eta) \\
	&+ i\, \mathcal{A}_{ijlm}\,\mathcal{H}_{GW}(k,\eta) \big]
	\, ,
\end{split}
\ee
where $S_{GW}$ and $\mathcal{A}_{GW}$ characterizes the amplitude and polarization of the GWs. With the above definition, we can write,
\beq
\delta({\bk}-{\bk}')\,S_{GW} & = & \frac{1}{(2\pi)^3}\mathcal{M}_{ijlm}\langle h_{ij}'({\bk})\, h_{lm}^{*'}({\bk}')\rangle \, \\
\delta({\bk}-{\bk}')\,\calH_{GW} & = & \frac{1}{(2\pi)^3}\mathcal{A}_{ijlm}\langle h_{ij}'({\bk})\, h_{lm}^{*'}({\bk}')\rangle. 
\label{GW-polEH}
\eeq
We now choose a coordinate system, for which unit vectors are $\hat{{\bf e}}_1$, 
$\hat{{\bf e}}_2$, $\hat{{\bf e}}_3$, in which GW  propagates in the $\hat{{\bf e}}_3$ 
direction. Further, we introduce 
\be
e_{ij}^\pm =-\sqrt{\frac{3}{2}} \,(\varepsilon_i^\pm\times  \varepsilon_j^\pm)
\ee
which forms basis for a tensor perturbations and satisfy 
the following properties:\\ $\delta_{ij}\,e^\pm_{ij}\,=\,0$, $\hat{k}_i\,e^\pm_{ij}\,=\,0$ and 
$e^\pm_{ij}\,e^\mp_{ij}\,=3/2$~\cite{wheeler:1973, Wayne:1997hm}. The right-handed and left 
handed circularly polarized state of the GWs are represented by $+$ and $-$ sign respectively.
In this basis, polarization tensor and modes of the GWs can be written as follows:
\beq
\Pi_{ij}({\bk}) & = & e_{ij}^+\Pi^+({\bk})+e_{ij}^-\Pi^-({\bk}) \label{eq:pol-pi_ij}\, , \\
h_{ij}({\bk}, \eta) & = & h^+({\bk}, \eta)e^+_{ij}+h^-({\bk}, \eta)e^-_{ij} \, .
\eeq
On using \eqn{eq:pol-pi_ij}, \eqn{f-g-function} can be expressed as
\be
\begin{split}
	\delta({\bsym k}-{\bsym k}')f(k) &=  \frac{3}{2}\langle \Pi^+({\bsym k})\Pi^{+*}({\bsym k}') +\Pi^-({\bsym k}) \Pi^{-*}({\bsym k'}) \rangle 
	\label{eq:f(k)}\, \\
\end{split}
\ee
\be
\begin{split}
	\delta({\bsym k}-{\bsym k}')\,g(k) &=  -\frac{3}{2}\langle \Pi^+({\bsym k})\Pi^{+*}({\bsym k}') -\Pi^-({\bsym k}) \Pi^{-*}({\bsym k'}) \rangle 
	\label{eq:g(k)} \, .
\end{split}
\ee
Adding and subtracting \eqn{eq:f(k)} and \eqn{eq:g(k)} we obtain,
\beq
f(k) + g(k) & = & 3\langle \Pi^-({\bk})\Pi^{-*}({\bk})\rangle\, \approx 3 \langle|\Pi^-|^2\rangle\, 
\label{eq:f+g}\\
f(k) - g(k) & = & 3\langle \Pi^+({\bk})\Pi^{+*}({\bk})\rangle \approx 3 \langle|\Pi^+|^2\rangle \, .
\label{eq:f-g}
\eeq
Similarly, we can write eq.(\ref{GW-polEH}) as:
\be
\small{
	\delta({\bsym k}-{\bsym k}')S_{GW}(k,\eta) = \frac{3}{2}
	\langle
	h^{+}({\bsym k},\eta)h^{+*}({\bsym k}',\eta)+
	h^{-}({\bsym k},\eta)h^{-*}({\bsym k}',\eta)
	\rangle}\nonumber
\ee
\be
\small{
	\delta({\bsym k}-{\bsym k}')\mathcal{H}_{GW}(k,\eta) =  -\frac{3}{2}
	\langle
	h^{+}({\bsym k},\eta)h^{+*}({\bsym k}',\eta)-
	h^{-}({\bsym k},\eta)h^{-*}({\bsym k}',\eta)
	\rangle} \nonumber
\ee
Therefore, components of the GWs evolve as
\be
h^{\pm ''}({\bk}, \eta)+ 2\,{a'\over a}\, h^{\pm '}({\bk}, \eta)+k^2 h^{\pm}({\bk}, \eta)=8\pi G \Pi^{\pm}({\bk})\, ,
\label{eq:hij-pm}
\ee
here $\Pi_{ij}$ is the mean square root value of the transverse traceless part of the energy momentum tensor. 
In terms of dimensionless variable $x=k\eta$, the above equation reduces to 
\be
h^{\pm ''} + 2 \frac{\alpha}{x}\, h^{\pm '} + h^{\pm} =  \frac{s^\pm}{k^2}\, ,
\label{eq:GW-source}
\ee
where $s^\pm(k,\eta) =  \lt({8\pi G\over a^2}\rt) \sqrt{\frac{f(k)\,\mp \,g(k)}{3}}$ and the parameters $\alpha=1$ and $\alpha=2$ indicates the radiation dominated and the matter dominated epoch respectively.  In the radiation dominated epoch, the homogeneous solution of the 
eq.~(\ref{eq:GW-source}) are 
the spherical Bessel function $j_0$ and $y_0$. In our case, magnetic field is generated 
at $\eta_{in}$ in the radiation dominated epoch due to chiral instability leading to 
anisotropic stress which in turn generates the gravitational waves. Thus, the general 
solution of \eqn{eq:GW-source} can be given as,
\be
h^\pm(x) = c_1^\pm(x)\,j_0(x)\,+\,c_2^\pm(x)\,y_0(x)
\label{eq:GW-solution}
\ee
where $c_1(x)$ and $c_2(x)$ are undetermined coefficients which is given as
\beq
c_1^\pm(x) & = & - \int_{x_{in}}^x  dx' \frac{s^\pm(x')}{{\rm w}(x')\, k^2}\,y_0(x') \label{eq:c1} \, \\
c_2^\pm(x) & = & \int_{x_{in}}^x  dx' \frac{s^\pm(x')}{{\rm w}(x')\, k^2}\, j_0(x') 
\label{c2}
\eeq
where ${\rm w}(x) = j_0(x) \, y_0'(x) - y_0(x) \, j_0'(x) = \frac{1}{x^2}$. We have calculated $c_1^\pm(x)$ and $c_2^\pm(x)$ using equations \eqn{eq:c1} and \eqn{c2} under the limits of $x>1$. 
In this limit, the second term with $y_0$ diverges, therefore, first term dominates over second one. In this case, in the radiation dominated epoch, the two polarizations of the tensor perturbations can be written as:
\be
h^+(x) =  c_1^+(1)j_0(x) 
=  -{90 \over \pi^2 g_{\rm eff}} \sqrt{\frac{f-g}{3}}\,j_0(x)\,{\rm log}(x_{in}) 
\label{eq:mode h+}
\ee
\be
h^-(x) =  c_1^-(1)j_0(x) \,  
=  -{90 \over \pi^2 g_{\rm eff}}
\sqrt{\frac{f+g}{3}}\,j_0(x)\,{\rm log}(x_{in})\, 
\label{eq:mode h-},
\ee
here $x=1$ in $c_1(1)$ signifies the value at the time of horizon crossing. After horizon crossing, these gravitational waves propagate without any hindrance. However, their energy and polarization stretched by the scale factor, similar to the magnetic radiation energy. The time derivative of the of the \eqn{eq:mode h+} and \eqn{eq:mode h-} is
\be
h^{+'}(x) 
=  -{90 \over \pi^2 g_{\rm eff}} \sqrt{\frac{f-g}{3}}\,j_0'(x)\,{\rm log}(x_{in}),
\ee
\be
h^{-'}(x)  
=  -{90 \over \pi^2 g_{\rm eff}} \sqrt{\frac{f+g}{3}}\,j_0'(x)\,{\rm log}(x_{in})\, ,
\ee
In real space, the energy density of the gravitational waves is defined as
\be
\rho_{GW} = {1\over 16\,\pi\, G\, a^2}\, \langle h'_{ij}h'_{ij}\rangle.
\ee
Note that a factor of $a^2$ in the denominator comes from the fact that $h'$ is the derivative with respect to conformal time. In Fourier space, the energy density of the gravitational wave is given as
\be
S_{GW}(k) =  \int \frac{dk}{k}\frac{d\,S_{GW}}{d\,{\rm log}\, k} \,
\ee
with
\begin{eqnarray}
\frac{d\,S_{GW}(k)}{d\, \text{log}k} &=&  \frac{k^3 }{2\,M_*^4\,a^2\,(2\pi)^6 G}(|h^{+'}|^2+|h^{-'}|^2) \, . 
\end{eqnarray}
With this definition, we can define power spectrum evaluated at the time of generation as
\begin{eqnarray}
\frac{d\Omega_{GW,s}}{d{\rm log}k}&=& \frac{1}{(\rho_{c,s}/M_*^4)}\frac{dS_{GW,s}}{d{\rm log}k}\\
&=&{16\pi k^3\over 3(2\pi)^6 a_s^2} \lt({90\over \pi^2g_{\rm eff}}\rt)^2{f(k)\over H_s^2}\lt[j_0'(x){\rm log}(x_{in})\rt]^2 . 
\nonumber 
\label{eq:GW-spectrum}
\end{eqnarray}
where $\rho_{c,s}$ is the critical density of the universe at the time of generation of GW.
Once gravitational waves are produced, they are decoupled from the rest of the Universe. This implies that the energy density of the gravitational waves will fall as $a^{-4}$ and frequency redshifts as $a^{-1}$. Hence, the power spectrum at today's epoch can be given as 
\be
\frac{d\Omega_{{GW,0}}}{d{\rm log}k}\equiv \frac{d\Omega_{GW,s}}{d{\rm log}k}\,\lt({a_s\over a_0}\rt)^4
\lt({\rho_{c,s}\over \rho_{c,0}}\rt)\, .
\label{eq:gw-spec-today}
\ee
Assuming that the Universe has expanded adiabatically which implies that the entropy per comoving volume is conserved leads to 
\begin{equation}
{a_s\over a_0} = \left({g_{\rm eff,0}\over g_{\rm eff,s}}\right)^{1/3}\left({T_0\over T_s}\right)\, ,
\label{eq:entropy-conse}
\end{equation}
where we have used $g_{\rm eff}$ for the effective degrees of freedom that contribute to the entropy density also. This is due to the fact that effective degrees of freedom that contribute to the energy and entropy densities are same at very high temperature. Therefore, Eq.(\ref{eq:gw-spec-today}), using eq. \eqref{eq:entropy-conse} reads as
\begin{eqnarray}
\frac{d\Omega_{{GW,0}}}{d{\rm log}k} & = & \lt({g_{\rm eff,0}\over g_{\rm eff,s}}\rt)^{4/3}\lt({T_0\over T_s}\rt)^4
\lt({H_s\over H_0}\rt)^2\,\frac{d\Omega_{{GW,s}}}{d{\rm log}k}\,  \nn \\
& = & {16\pi k^3\over 3(2\pi)^6 a_s^2}\times\lt({90\over \pi^2g_{\rm eff}}\rt)^2\lt({g_{\rm eff,0}\over g_{\rm eff,s}}\rt)^{4/3} \nn \\
& & \lt({T_0\over T_s}\rt)^4 {f(k)\over H_0^2}\lt[j_0'(x){\rm log}(x_{in})\rt]^2
\end{eqnarray}
In figures (\ref{fig:chgw_k_diff_T}) and (\ref{fig:chgw_k_diff_N}), we have shown the variation of GW spectrum with respect to $k$ for different temperature at fix number density and for different number density at fix temperature. 
\section{Results and Discussion}
\label{sec:results}
Before discussing the results obtained, we would like to explain various important length scales useful for the magnetohydrodynamic discussion of the generation of GWs due to the chiral instability in presence of the external magnetic fields. Firstly, magnetic modes grow exponentially for $k= k_{{\rm ins}}= \xi^{(B)}/2 \approx g^2 (\frac{\mu}{T})T$ \cite{Anand:2017zpg, Bhatt:2015ewa}. Secondly, dissipation due to the finite resistivity of the plasma works at wave numbers $k< k_{{\rm diss}}\sim \sqrt{\frac{g_{\rm eff}}{6}}gT$ \cite{Weldon:1982ha}. Therefore, near instability, the wavenumber corresponding to the length scales of interest $k \ll k_{{\rm diss}}$. In the present analysis, we have not considered any dissipation in the plasma and restricted ourself to the scales where there is maximally growing modes of the magnetic fields are available. Therefore, for  $k$ values larger than the instability scale, this analysis may not be reliable. In figure (\ref{fig:chgw_k_diff_T}) and (\ref{fig:chgw_k_diff_N}), we have found that GWs peak occurs at $k_{{\rm ins}}$. It is evident that at higher values of $k$ i.e. at small length scales, power increases after instability. This is related to the rise in the magnetic energy at large $k$. The rise in the magnetic energy is unphysical as we know that in turbulent system, energy accumulates at smallest scales. This effect in principle can be restricted by going to hyper diffusion scale (instead of $\nabla^2$ operator, one needs to introduce $\nabla^4$ operator) \cite{Brandenburg:1996fc}.
We would like to emphasize, in figures (\ref{fig:chgw_k_diff_T}) and (\ref{fig:chgw_k_diff_N}) that variable $x=k\eta$  is a dimensionless quantity. In order to interpret the results, we convert $x$ in frequency of the GW. Since, $x=a\, k\eta = \frac{k}{T}\eta~= \frac{2\pi \nu}{T}\eta$. 
From this we can get $\nu$ in terms of $x$ as $\nu= \frac{T}{2\pi \eta} x$. Moreover, peak of the power spectrum of the GW occurs when growth of the instability is maximum which is given by $\nu_{{\rm max}}\approx \frac{16}{9\pi}\left(\frac{\delta^2}{\sigma/T}\right) T$ \cite{Bhatt:2015ewa, Weldon:1982ha}. Here $\delta$ is defined as $\delta= \alpha ( \mu_R-\mu_L) /T$. The red-shifted value of the frequency can be obtained using the relation: $\nu_0=\frac{T_0}{T_*}\nu_{{\rm max}}$, where $T_*$ is the temperature at which instability occurs. Hence, we can obtain the frequency at which maximum power is transferred from the magnetic field to GWs. The obtained formula of the frequency in simplified form is $\nu \approx 10^{9} \, \delta^2$ Hz, where we have used $T_0=2.73$ K $\approx 10^{11}$ Hz in our units and $\sigma/T=100$.  For temperature $T\sim 10^6$ GeV, $( \mu_R-\mu_L)/T \sim  10^{-3}$ \cite{Tashiro:2012mf} and thus, $\delta^2= 10^{-10}$ (with $\alpha\approx 10^{-2}$). Hence frequency where maximum power of GW occurs, is around $10^{-1}$ Hz. Thus they may be detected in eLISA experiment \cite{Moore:2014lga}.  Further, the strength of magnetic field changes when chiral charge density $n$ change. Fig. (\ref{fig:chgw_k_diff_N}) shows the effect of $n$ on GW spectrum. It is apparent that the $k_{{\rm ins}}$ is not affected by the number density and hence, the peak does not shift. However, the power in a particular $k$ mode enhances with an increase in $n$. This happens due to the fact that for a larger value of $n$, magnetic field strength is 
higher at larger $k$~\cite{Anand:2017zpg}.
\section{Conclusion}\label{sec:conc}
In the present work, we have extended our earlier works on the generation of primordial magnetic fields in a chiral plasma~\cite{Bhatt:2015ewa, Anand:2017zpg} to the generation of GWs. This kind of source may exist much above electroweak scale. We have shown that the gravitational anomaly generates the seed magnetic field which evolves and create  instability in the system. This instability acts as a source of anisotropic stress which leads to the production of gravitational waves. The production and evolution of the magnetic field has been studied using ~\eqn{eq:magevol}. In order to obtain the velocity profile, we have used scaling properties \cite{Olesen:1996ts,Yamamoto:2016xtu} rather than solving the Navier- Stokes equation. This scaling property results in more power in the magnetic field at smaller $k$ as compared to that of the case without scaling symmetries (see~\cite{Anand:2017zpg}). We have calculated power spectrum of the produced GWs and shown that the spectrum has a distinct peak at $k_{{\rm ins}}$ and hence correspond to the dominant frequency of GW. The GW generated at high temperature $T \geq 10^{6}$ GeV via aforementioned method is potentially detectable in eLISA. 

In this work, we have considered massless electrons much above electroweak scale and discussed the production of gravitational waves due to chiral instabilities in presence of Abelian fields belonging to $U(1)_Y$ group. However, a similar situation can arise in the case of Quark-Gluon Plasma (QGP) at $T\gtrsim 100$ MeV where quarks are not confined and interact with gluons which may result in instabilities. Thus, GW can be produced in QGP as well. 

To conclude, the study of relic GWs can open the door to explore energy scales beyond our current accessibility and give insight into exotic physics.
\section*{Acknowledgments}
We would like to acknowledge Late Prof. P. K. Kaw for his insight and motivation towards the problem. We also thank Abhishek Atreya for discussions.
\bibliographystyle{apsrev4-1} 
\bibliography{ref} 
\end{document}